\documentclass[preprint2]{aastex}          %% The manuscript based on AASTeX v5.x
\usepackage{spr-astr-addons}    %% mimicing ASTR journal style
\usepackage{url}

\usepackage[utf8]{inputenc}
\usepackage[T1]{fontenc}
\usepackage{amssymb}

\begin{document}

%% Article title
%
\title{Are black holes in an ekpyrotic phase possible?}

%% Running heads
\shorttitle{Are black holes in an ekpyrotic phase possible?}
\shortauthors{J. C. S. Neves}

%% Author and Affilations
\author{J. C. S. Neves\altaffilmark{*}}
\affil{Instituto de Matemática, Estatística e Computação Científica, Universidade
Estadual de Campinas,  CEP. 13083-859, Campinas-SP, Brazil}

%% Alternate Affilations
\altaffiltext{*}{E-mail: nevesjcs@ime.unicamp.br}
%\altaffiltext{3}{}

%% Abstract
\begin{abstract}
The ekpyrotic phase (a slow contraction cosmic phase before the current expansion phase) manages to solve the main problems of the standard cosmology by means of a scalar field interpreted as a cosmic fluid in the Friedmann equation. Moreover, this phase generates a nearly scale-invariant spectrum of perturbations in agreement with the latest data. Then, the ekpyrotic
mechanism is a serious possibility to the inflationary model. In this work—by using the approach of deforming metrics at linear level—we point out that it is impossible to generate a black hole with spherical symmetry supported by an isotropic fluid in this cosmological scenario.

\end{abstract}

%% Keywords
\keywords{Black holes, Cosmology}

%%  Please use labels (\label, \ref) for section, figure, table, 
%%  equation  reference. Use \cite for bibliography references.
%
\section{Introduction}%\label{s:?}

The inflationary mechanism is still the most popular alternative to
solve the main problems of the standard cosmology: the flatness, homogeneity,
horizon and isotropy problems. But this mechanism is not the only
one. The ekpyrotic mechanism manages to solve these problems by means
of a previous phase of contraction before the current phase of expansion
(see \cite{Lehners} for a review). In this slow contraction phase,
the universe is dominated by a stiff fluid with a determined equation of
state (EoS): $w\equiv\frac{p}{\rho}>1$ ($p$ and $\rho$ are pressure
and energy density, respectively). With this fluid, the universe became
flat, homogeneous and isotropic before the expansion phase, and the
seeds of the structure formation were generated during this contraction
phase. These seeds, generated by quantum fluctuations, are nearly
scale-invariant and are in agreement with the latest data \citep{Collaborations,Collaborations2,Collaborations3,Collaborations4,Collaborations5}. 

Among the researches of the ekpyrotic universe, there exists the belief
that this mechanism solves the problems of the standard cosmology
as well as avoids the problems of the inflationary paradigm \citep{Steinhardt}.
According to these researches, the inflationary mechanism has two
important problems: the initial conditions and multiverse-unpredictability.
On the other hand, the ekpyrotic cosmology does not have both problems
and\textemdash{}with the aid of other ingredients, for example, the Galileon
cosmology and ghost condensate \citep{Various-Galileon-Ghost1,Various-Galileon-Ghost2,Various-Galileon-Ghost3}\textemdash{}manages to avoid another issue in standard cosmology: the problem of the initial
singularity. As we can see, the ekpyrotic cosmology offers us an interesting
option to the inflation.

Moreover, the ekpyrotic cosmology may be extended to a cyclic cosmology. In this cyclic scenario \citep{Steinhardt2}, one has a endless sequence where the universe expands and contracts. There is no beginning for the time\textemdash{}therefore, the problem of initial conditions is solved in this extended version. The ekpyrotic mechanism is able to remove debris generated in a previous cycle, such as black holes (BHs). Thus the existence of such a mechanism is crucial to build cyclic cosmologies. 

Any modern scientific cosmology has the isotropy as a feature. From the beginning of the modernity, the cosmology assumes that there is no favourite place or direction in the universe (contrary to both the Plato's cosmology in his \textit{Timaeus} and the Aristotle's view in \textit{On the Heavens}). The Physics, from this time, describes the universe as isotropic at large scales. Hence, the current scientific cosmology, an Einsteinian cosmology, is assumed to be isotropic. 

In this work, by using a determined approach, we intend to show that BHs supported by
an isotropic fluid exclude an EoS given by the ekpyrotic phase. 
From the mechanism developed in \cite{Various}, \cite{Various2,Various3}, \cite{Various4}, \cite{Various5}, \cite{Various6}, \cite{Molina2} and generalized in \cite{Molina}, we study, using the Schwarzschild metric as ansatz, the
family of deformed BHs with spherical symmetry supported by an isotropic
fluid obtained in \cite{Molina2}. We point out that these solutions are
forbidden when supported by a stiff fluid ($w>1$). This result may be interpreted as an indirect indication of the ekpyrotic
phase as a period that removes inhomogeneities, anisotropies and debris of a previous cosmic phase in cyclic cosmologies that use the ekpyrotic mechanism. 

The structure of this paper is presented in the following: in Section
2 we presented the ekpyrotic mechanism and its main features; in
Section 3 we show the solutions, supported by a isotropic fluid,
constructed by means of deformations in the Schwarzschild solution;
in Section 4, the final remarks. We adopt the metric signature $diag(-+++)$
and $G=c=1$, where $G$ is the Newtonian constant, and $c$ is the speed of light in vacuum. 

\section{The ekpyrotic phase}

Such as the standard Big Bang cosmology, the ekpyrotic cosmology assumes a Friedmann-Lemaître-Robertson-Walker (FLRW) metric as a solution of the gravitational field equations to describe the spacetime fabric. In the first version \cite{Khoury},
a brane world model was used. But, some years later, a four-dimensional version, an effective model, was introduced in \cite{Buchbinder}. Assuming the FLRW metric
\begin{equation}
ds^{2}=-dt^{2}+a(t)^{2}\left[\frac{dr^{2}}{1-kr^{2}}+r^{2}\left(d\theta^{2}+sin^{2}\theta d\phi^{2}\right)\right],\label{FRLW}
\end{equation}
where $a(t)$ is the scale factor, which depends on the cosmic time
$t$, $k$ determines the spatial curvature ($k=0,$ flat universe;
$k=-1,$ open universe; $k=1,$ closed universe), and the cosmic matter-energy
content is described by a perfect fluid $T_{\nu}^{\mu}=diag(-\rho,p,p,p)$, the generalized Friedmann equation reads
\begin{equation}
H^{2}=\frac{1}{3}\left(-\frac{3k}{a^{2}}+\frac{\rho_{m}}{a^{3}}+\frac{\rho_{r}}{a^{4}}+\frac{\sigma^{2}}{a^{6}}+...+\frac{\rho_{\phi}}{a^{3(1+w_{\phi})}}\right),\label{Friedmann}
\end{equation}
with $H=\dot{a}/a$, where dot represents a derivative w.r.t. time $t$; $\rho_{m}$ and $\rho_{r}$ mean the energy densities of non-relativistic matter (including dark matter) and radiation, respectively. The term which depends
on the $a^{-6}$ indicates the energy density of anisotropies. Lastly,
$\rho_{\phi}$ denotes the energy density of the scalar field that
generates the ekpyrotic phase—interpreted as a perfect
fluid in the Friedmann equation. Such a field has the action
\begin{equation}
S_{\phi}=\int d^{4}x\sqrt{-g}\left(-\frac{1}{2}g^{\mu\nu}\partial_{\mu}\phi\partial_{\nu}\phi-V(\phi)\right),\label{action}
\end{equation}
where $g$ is the metric determinant, and an EoS written as 
\begin{equation}
w_{\phi}=\frac{p_{\phi}}{\rho_{\phi}}=\frac{\frac{1}{2}\dot{\phi}^{2}-V(\phi)}{\frac{1}{2}\dot{\phi}^{2}+V(\phi)}.\label{EoS_ekpyrotic}
\end{equation}
Typically, the potential is given by $V(\phi)=-V_{0}e^{-c\phi}$,
with $V_{0}$ and $c$ playing the role of constants. A very steep
potential of this type produces an EoS for the ekpyrotic phase with a dominant fluid given
by $w_{\phi}>1$. Such a matter content is dominant in Eq. (\ref{Friedmann})
during a contraction phase ($a^{-n},$ with $n>6$). The term that
describes the anisotropies, $\sigma^{2}/a^{6},$ is not dominant over
the last term in Eq. (\ref{Friedmann}). Then, the slow contraction phase,
where the stiff fluid is dominant over other matter contents,
produces a homogeneous and isotropic universe, free of the Belinsky-Khalatnikov-Lifshitz (BKL) instabilities, before the current
expansion phase. The authors in \cite{Garfinkle} have shown that the ekpyrotic mechanism works even when the initial conditions of this slow contraction phase are highly inhomogeneous and anisotropic. 

However, the standard ekpyrotic mechanism with one scalar field is considered tachyonically unstable (see, for example, \cite{Mingzhe}). To solve this problem, another scalar field is introduced to generate a stable mechanism with nearly scale-invariant perturbations and small non-Gaussianity \citep{Ijjas}.

In the next section, we will show how to build BH solutions supported by an isotropic fluid. To obtain these BHs, the approach which deforms BHs will be presented. Contrary to the work \cite{Garfinkle}, in this work, we do not study the initial conditions of the ekpyrotic phase and its evolution. In this article, we assume an isotropic fluid (as a linear constraint in the deformation approach) and show that an ekpyrotic phase does not support these deformed BHs. That is, an isotropic universe with EoS $w>1$, or an ekpyrotic phase where the universe is already isotropic (close to the bounce), does not support the deformed geometries. The novel part of this work is the use of the deformation approach to show these results.         

\section{Constructing isotropic black holes}

It is possible to build isotropic BH solutions by means of deformations. From the standard singular solutions, such as Schwarzschild, Reissner-Nordström, Schwarzschild-(A)-dS and others \citep{Various,Various2,Various3,Various4,Various5,Various6,Molina2}, and regular solutions \citep{Neves,Neves2}, at linear level, deformed BHs are obtained by imposing a linear constraint in the energy-momentum tensor (the approach was generalized in \cite{Molina}, where other constraints were used). These deformed solutions are close to the standards BHs. A linear constraint in the energy-momentum tensor to generate isotropic solutions means:
\begin{equation}
p_{r}=p_{t},\label{constraint_2}
\end{equation}
where
\begin{equation}
R_{\nu}^{\mu}-\frac{1}{2}R\delta_{\nu}^{\mu}=8\pi T_{\nu}^{\mu}=8\pi\left(\begin{array}{cccc}
-\rho\\
 & p_{r}\\
 &  & p_{t}\\
 &  &  & p_{t}
\end{array}\right).\label{field_equations}
\end{equation}
$p_{r}$ and $p_{t}$ are the radial and tangential pressures, respectively, and assume the same value due to constraint (\ref{constraint_2}) throughout this work. 
The constraint (\ref{constraint_2}) used in the field equations (\ref{field_equations}),
with the aid of the general metric with spherical symmetry
\begin{equation}
ds^{2}=-A(r)dt^{2}+\frac{1}{B(r)}dr^{2}+r^{2}\left(d\theta^{2}+sin^{2}\theta d\phi^{2}\right),\label{metric}
\end{equation}
yields to
\[
-\frac{1}{r^{2}}\left[1-B(r)\right]-\frac{1}{2}\left[\frac{A'(r)B'(r)}{A(r)}-\left(\frac{A'(r)}{A(r)}\right)^{2}\right]
\]
\begin{equation}
+\frac{1}{2r}\left[\frac{B(r)A'(r)}{A(r)}\right]-\frac{1}{2r}B'(r)-\frac{A''(r)B(r)}{2A(r)}=0.\label{constraint_3}
\end{equation}
 It is worth noting that the deformed solution has $A(r)\neq B(r).$
In this sense, the Birkhoff theorem is not valid. 

To obtain a set of solutions close to the known solutions, we choose
a function $A(r)=A_{0}(r)$, and the approach yields to a function
$B(r)$ close to the standard case where $B(r)=A(r)$ in Eq. (\ref{metric}). The approach modifies only the function $B(r)$ because this modification leads to the solutions with a different spacetime structure (event horizons, as we will see below).
In this work, we follow \cite{Molina2} and deal with deformations
from the Schwarzschild metric, where
\begin{equation}
A(r)=A_{0}(r)=1-\frac{2m}{r}.\label{A_0}
\end{equation}
The ansatz is completed by
\begin{equation}
B(r)=A_{0}(r)+\left(C-1\right)B_{lin}(r).\label{B}
\end{equation}
The subscript $lin$ stands for linear, and $C-1$ may be seen as the first-order expansion coefficient. Then, we want to obtain deformed solutions until the linear level (in \cite{Molina2} there is a discussion on the possibility of generating deformed solutions beyond the linear level). According to \cite{Molina}, the constant $C$ fixes basically three kind of solutions: for $C>0$ one has either regular or singular black holes, $C<0$ may determine wormholes, and for $C=C_{0}$, one has extremal black holes. For $C=1$ the standard cases are restored. 

In the linear regime, the function $B_{lin}(r)$ is determined from
\begin{equation}
B_{lin}(r)=A_{0}(r)\ exp\left[-\int\frac{dr}{rh(r)}\right],\label{Blin2}
\end{equation}
which is solution of Eq. (\ref{constraint_3}) using the ansatz (\ref{A_0})-(\ref{B}),
where 
\begin{equation}
h(r)=-\frac{1}{2}A_{0}(r)-\frac{1}{4}rA_{0}'(r).\label{h(r)}
\end{equation}
The zeros of $h(r)$ are important because may determine the singularities
of the spacetime. In this work, $h(r)$ has one real root, $r_{0}= m$. Using the ansatz (\ref{A_0}) in the Eqs. (\ref{Blin2})-(\ref{h(r)}),
one has
\begin{equation}
B(r)=A_{0}(r)\left[1+\left(C-1\right)\left(r-m\right)^{2}\right],\label{Blin3}
\end{equation}
 and the metric (\ref{metric}) is completely determined. The zeros of $B(r)=0$ fix the horizons (when a metric is written in the form (\ref{metric}), the localization of horizons is given by the zeros of $g^{rr}=B(r)=0$). According
to \cite{Molina2}, BHs are possible only with $C>0$ (our value of
$C$ corresponds to the $C/m^{2}+1$ in the cited paper). 

For $C>1$, the function $B(r)$ has two zeros, $r_{-}<r_{+}$. The
largest zero corresponds to the event horizon $r_{+}=2m$. The metric
is singular at $r=r_{0}<r_{-}$. Then, the spacetime  structure reads
\begin{equation}
r_{0}<r_{-}<r_{+}<\infty.\label{structure_1}
\end{equation}
This case corresponds to a singular BH in a noncompact universe. 

When $0<C<1$, the situation is quite different: there are three zeros
in $B(r)$ (the third zero is not zero of $A(r)$). There is a maximum
value of $r$ $(r_{max}>r_{+})$. According to \cite{Molina2}, these
values of $C$ lead to a singular BH in a compact universe. The spacetime
 structure is given by
\begin{equation}
r_{0}<r_{-}<r_{+}<r_{max},\label{structure_2}
\end{equation}
 with
\begin{equation}
\lim_{C\rightarrow1}r_{max}=\infty.\label{rmax}
\end{equation}
 Then, these are the two types of solutions when the ansatz is given
by Eqs. (\ref{A_0})-(\ref{B}) assuming isotropy. 

With $A(r)$ and $B(r)$ fixed, the values of the components of the
energy-momentum tensor are:
\begin{eqnarray}
\rho & = & -\frac{\left(C-1\right)\left(r-m\right)\left(3r-5m\right)}{r^{2}},\label{energy_density}\\
p_{r} & = & p_{t}=p=\frac{\left(C-1\right)\left(r-m\right)^{2}}{r^{2}}.\label{pressure}
\end{eqnarray}
With these components in the energy-momentum tensor, an EoS is possible:
\begin{equation}
w=\frac{p}{\rho}=-\frac{r-m}{3r-5m},\label{w}
\end{equation}
which is almost constant and always negative outside the event horizon (see Fig. 1). For large $r$, $w$ is about $-1/3$. In the case $C>1$, one has the simple limit 
\begin{equation}
\lim_{r\rightarrow\infty}w=-\frac{1}{3}.\label{lim}
\end{equation}
At the event horizon, one has $w(r_{+})=-1$. In this sense, these deformed metrics may not be supported (such as the Schwarzschild-(A)-dS metric is by a fluid with EoS $w=-1$) by a stiff fluid  ($w>1$), dominant in the ekpyrotic phase. Then, at linear level, BHs supported by an isotropic stiff fluid are ruled out. But not only a stiff fluid: as we can see, these metrics are not supported by either a fluid with EoS $w=1/3$ (radiation)
or a phantom-like fluid ($w<-1$). Outside the event horizon, the EoS assumes $-1<w<-1/3$, which are values corresponding to a quintessence fluid. 

\begin{figure}[tb]
\begin{centering}
\includegraphics[scale=0.38]{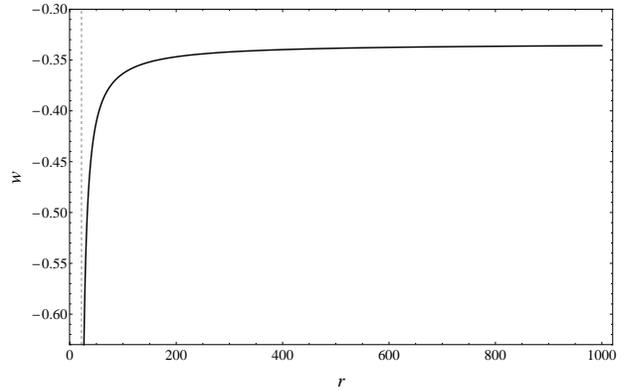}
\par\end{centering}

\caption{$w=\frac{p}{\rho}$ obtained from isotropic deformed solutions. As
we can see, this EoS is pretty much constant outside the event horizon
and rules out a stiff fluid $(w>1)$, dominant in the ekpyrotic phase. The dashed line indicates
the event horizon. In this graphic we use $m=11$.}
\end{figure}

\section{Final remarks }

The ekpyrotic cosmology is an option for the inflationary mechanism
to solve the problems of the standard cosmology (including the initial
singularity) and to generate nearly scale-invariant quantum fluctuations.
This alternative cosmology is characterized by a slow phase of contraction, where the equation of
state (EoS) of the matter-energy contend in this phase is dominated by a stiff fluid $(w\equiv\frac{p}{\rho}>1)$.

We show that—by means of a procedure of generating deformed isotropic black holes (BHs) at linear level in $B(r)$, adopting a linear constraint in the energy-momentum tensor—the metrics with spherical symmetry obtained from the Schwarzschild metric are not supported by a stiff fluid. These metrics are not supported by a stiff fluid in the same sense that the Schwarzschild-(A)-dS metrics may be interpreted as solutions with spherical symmetry supported by a fluid with EoS $w=-1$. We believe that this result may be seen as an \textit{indirect} indication (it is not a strong proof) of the ability of the ekpyrotic phase to leave the universe free of inhomogeneities and debris (and BHs may be interpreted as debris) in a previous phase before the current phase of cosmic expansion. This result may be important for cosmological cyclic models that assume an ekpyrotic phase. 

\acknowledgments
This work was supported by Fundação de Amparo à Pesquisa do Estado de São Paulo (FAPESP), Brazil (Grant No. 2013/03798-3). I would like to thank Alberto Saa for comments and suggestions.

\end{document}